\documentclass[showpacs,twocolumn,prl]{revtex4}
\usepackage{graphicx}
\usepackage{epsfig}
\usepackage{amssymb}
\usepackage{color}

\def\beq{\begin{equation}}
\def\eeq{\end{equation}}
\def\beqa{\begin{eqnarray}}
\def\eeqa{\end{eqnarray}}

\def\e{\epsilon}

\def\D{\Delta}
\def\del{\delta}
\def\e{\varepsilon}
\def\cH{{\mathcal H}}

\def\etal{{\sl et al.}}
\def\o{\omega}

\def\nonum{\nonumber \\}

\def\del{\delta}

\def\dag{\dagger}

\def\nonum{ \nonumber \\}

\def\urusi{URu$_2$Si$_2$}
\def\Tho{T_{\mathrm{HO}}}

\begin{document}

\title{Hybridization wave as the 'Hidden Order' in   \urusi}
\author{Yonatan Dubi$^1$\footnote{currently at the school of physics and astronomy, Tel-Aviv University, Israel} and Alexander V. Balatsky$^{1,2}$}
%\author{ DRAFT ~~DRAFT ~~DRAFT~~ // please do not circulate}
\affiliation{$^1$ Theoretical Division, Los Alamos National Laboratory, Los Alamos, NM 87545, USA}
\affiliation{$^2$ Center for Integrated Nanotechnologies, Los Alamos National Laboratory, Los Alamos, NM 87545}
%\pacs{74.78.-w,74.20.-z,74.40.+k}
\begin{abstract}
 A phenomenological model for the 'hidden order' transition in the heavy Fermion material \urusi ~is introduced. The 'hidden order' is identified as an incommensurate, momentum-carrying hybridization between the light hole band and the heavy electron band. This modulated hybridization appears after a Fano hybridization at higher temperatures takes place. We focus on the  hybridization wave  as the order parameter in \urusi ~
  and possibly other materials with similar band structures. The model is qualitatively  consistent with numerous experimental results obtained from e.g. neutron scattering  and scanning tunneling microscopy. Specifically, we find a  gap-like feature in the density of states and the appearance of features at an incommensurate vector $Q^*\sim 0.6 \pi/a_0$. Finally, the model allows us to make various predictions which are amenable to current experiments.    \end{abstract}
\maketitle

 The heavy Fermion material \urusi~ exhibits an enigmatic 'hidden order' (HO) transition at a temperature $\Tho
=17.5$K \cite{Palstra,Schlabitz,Maple}. Although being more than two decades old, the nature of the transition (which is a typical mean-field second order transition) and the order parameter which describe it are still unknown
\cite{Tripathi,Shah}. Various theories have been put forward to explain the transition, some of them describing it as a localized one \cite{Chandra,Santini,Kasuya,Sikkema,Cricchio,Haule}
%(perhaps motivated by the large entropy release observed in the transition, see discussion in the supplementary material),
and some describe it as an itinerant phenomenon \cite{Ikeda,Varma,Elgazzar,Harima,Balatsky}.
% Here we propose a theory for the HO in \urusi, which is in accord with various recent experimental findings: the gap-like suppression of electronic states, the Fano lattice structure,  the quasiparticle interference as seen by scanning tunneling microscopy, as well as features appearing in Neutron scattering
%experiments.  We identify the order parameter as an incommensurate hybridization between the light and heavy fermion bands which lie close to the Fermi energy. An essential ingredient is that the HO appears in the presence of an
%existing Fano coupling between the light band and the localized $f$-electrons, giving rise to a unique tunneling
%line-shape. Detailed predictions of the theory can be tested within current experimental techniques.

%The importance of solving this puzzle lies not only in its academic interest, but also in our ability to design materials with specific functionality, which is one of the ultimate goals of research in heavy Fermion materials, and is one of the reasons for which new theoretical and experimental methods have been devised in an attempt to resolve the puzzle.

In the last few years various new experimental techniques, previously unavailable, were presented, and now provide additional valuable hints as to the origin of the Hidden Order transition. The
key experimental observations relevant for our discussion are:

(i) In inelastic neutron-scattering measurements \cite{Wiebe} (as well as other measurements \cite{Broholm}) enhanced
scattering is observed at the incommensurate wave vectors $Q^* \sim 0.6, 1.4~ \pi/a_0$ below the HO transition. A
spin-gap-like feature of $\D \sim 5 meV$ is observed, as extracted from the heat capacity measurements. In many heavy-Fermion systems the spin-gap is indicative of the accompanying charge gap \cite{Riseborough}.

(ii) Angle-resolved photoemission spectroscopy measurements indicate that the heavy $f$-band is very weakly
dispersive, crosses the Fermi energy at the HO transition, and settles slightly below it \cite{Santander}.

\begin{figure}
\includegraphics[width=8truecm]{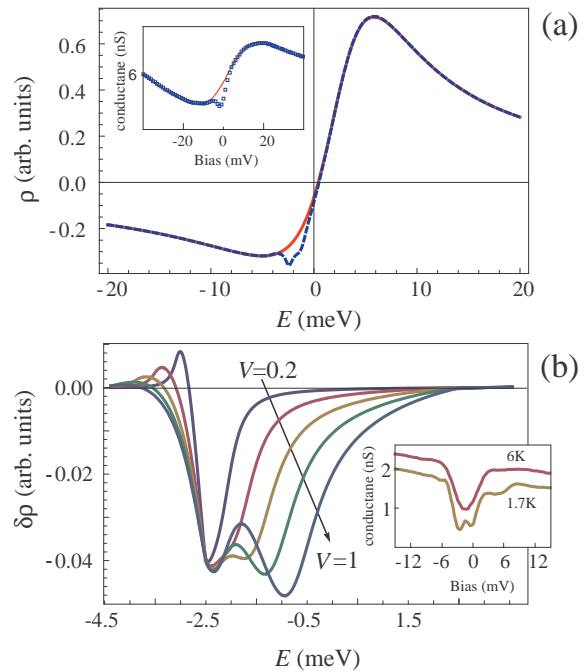}
\caption{ (a) Local DOS for a $d$-band
coupled to the localized $f$-electrons, exhibiting an asymmetric Fano line-Shape (solid line, see text for
parameters). The dashed line is the local DOS in the presence of the hidden order. These are in good agreement with
the DOS (proportional to conductance) seen in STM experiments, shown in the inset (extracted with permission from
\cite{Schmidt}). (b) Change in DOS due to the HO coupling, $\del \rho=\rho(V)-\rho(V=0)$, for $V=0.2,0.4,...,1$. The
main features are the appearance of a gap structure (which scales with $V$, see Fig.~\ref{Gap}), and a peak structure
within the gap, appearing for large enough $V$ (corresponding to low enough temperature). Both these features have
been observed experimentally \cite{Schmidt,Aynajian}, as seen in the inset (extracted with permission from
\cite{Schmidt}).}\label{DOS}
\end{figure}

(iii)  A Fano line-shape in
the density of states (DOS) which starts to form at $\sim120$K has been seen in recent scanning tunneling microscopy (STM) measurements \cite{Schmidt,Aynajian}. This Fano line-shape is fully developed at $\sim 20$K (inset of
Fig.~\ref{DOS}(a)).

 (iv) At the HO transition temperature, the bottom of the Fano line-shape develops a gap-like feature which evolves
 with temperature \cite{Schmidt,Aynajian} (inset of Fig.~\ref{DOS}(b)). Both the Fano parameters and the gap
 structure depend on the STM tip position and modulate in space with the positions of the different atoms.

(v) The hole band develops a hybridization feature below the HO transition, corresponding to momentum $Q\approx 0.3~
\pi/a_0$, close to where the heavy Fermion band crosses it \cite{Schmidt}.

\begin{figure}
\includegraphics[width=6.5truecm]{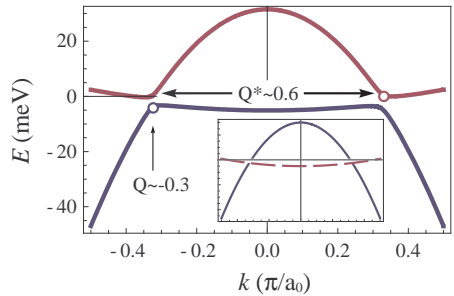}
\caption{Schematic representation of the band structure assumed in the
model and the incommensurate scattering. The Heavy electron $f$-band and the light hole $d$-band cross at $Q\approx
\pm 0.3$ (inset). The resulting scattering hybridizes electrons with momentum $Q$ and holes with momentum $-Q$ (and
vice versa), generating a hybridization wave which is the HO order parameter. The $Q^*=2Q\approx 0.6$ momentum
transfer is in accord with Neutron scattering experiments \cite{Wiebe}.}\label{bands}
\end{figure}

 These observations seem to suggest that a description of the HO in terms of reorganization of the localized degrees of freedom may be incomplete. Guided by the experimental findings, we construct a phenomenological theory for the HO transition. A one-dimensional model system is taken for simplicity, which contains a light hole ($d$) band and a heavy electron ($f$-electrons, dashed line) band, depicted in the inset of Fig.~\ref{bands}. We also performed the same calculation using bands generated by local density approximation (LDA) as input (to be shown elsewhere).
  Both from the model and from LDA based bands we take  the bands crossing  at momentum points $Q \approx \pm 0.3$ (
  along (1,0), (0,1) in units of $\pi/a_0$, where $a_0$ is the lattice spacing, in which all momenta are
  measured hereafter). The holes first hybridize with the local part of the $f$-electrons, giving rise to a Fano line-shape in
  the DOS \cite{Madhavan,Yang,Maltseva,Figgins,Haule} at temperatures that are much higher then the HO transition.  The band-structure gives rise to enhanced hybridization between electrons with momentum $Q$ and holes with momentum $-Q$ (and vice versa, Fig.~\ref{bands}) at specific points along the Fermi surface ("hot spots"). The resulting electron-hole coherence (which may be considered as an {\em indirect exciton}) is the HO order parameter. Since the "hot spots" are located at the band crossing, this incommensurate scattering generates a gap
  structure in the DOS with a line-shape similar to that observed in STM experiments \cite{Schmidt,Aynajian}.

The starting point of the model is a two-band hamiltonian,
\beq \cH=\sum_k \e^{(c)}_k c^\dagger_{k} c_k+\sum_k \e^{(f)}_k f^\dagger_{k}f_k \label{H0}~~,\eeq where $c^\dagger_k$
($f^\dagger_k$) creates a hole (electron) with momentum $k$. For simplicity (and as
it plays no important role in the physics we describe) the electronic spin is omitted. In real materials the  $d-f$ hybridization sets in at much higher
temperatures then HO transition and hence the remnant heavy band and light hole bands we use here are already
hybridized. We assume that lighter band is more $d$-character and heavier band has more $f$-character near the
$\Gamma$ point where we concentrate.

The origin of the Fano lineshape has been a subject of recent study \cite{Figgins,Maltseva,Yang,Fogelstrom}. In the absence of either translation invariance breaking or level broadening, a Fano lineshape should not appear \cite{Wolfle}.
A simple mechanism for its appearance is based on local scattering between the $d$-electrons and the $f$-electrons under the STM tip (the effect of the tip may not be negligible, \cite{Cheng}). The hamiltonian is then given by $ \cH_F=V_0 \sum_{k} c^\dagger_k f_{r=0}+h.c. ~~, $  where $f_{r=0}$
creates an $f$-electron on the lattice site ${r=0}$, i.e. under the tip. Alternatively, a similar result as below can be obtained by assuming full decoherence between the $f$-orbitals which lie on different lattice sites \cite{Wolfle}.

%To obtain the local $d$-band DOS, we write down the Dyson equation for the Green's function of the $d$-electron and
%the $f$-electron (for the latter, the only change is in the $k=0$ part of the Green function),
%\beqa
%g_{k,k'}&=& g^{(0)}_k \del_{k,k'}+V^2_0 g^{(0)}_k f_0 \sum_{q} g_{q,k'} \nonum
%f_0&=&f^{(0)}_0+V^2_0 \sum_{q,q'} g_{q,q'} f_0 ~~,\label{Dyson1} \eeqa
%where $g^{(0)}_k$ and $f^{(0)}_0$ are the bare Green's functions and $\del_{k,k'}$ is the Kroncker delta function.
%Solving the Dyson equation (and maintaining second order in $V_0$ for the denominator of $g_{k,k'}$)

It is now straight forward to write the Dyson's equations, with the solutions for the Green's functions for the $f$- and $d$-electrons
$f = \frac{1}{\o-\e_0-V^2_0 \chi_0},~~
g_{k,k'} =  g^{(0)}_k \del_{k,k'} +\frac{V^2_0}{\o-\e_0-V^2_0 \chi_0} g^{(0)}_k g^{(0)}_{k'}~~.$ Here $g_k^{(0)}$ is the bare d-electron propagator, and its susceptibility $\chi_0=\sum_k g^{(0)}_k =- \Gamma_0 (i-q)$, which defines $\Gamma_0$ (proportional to the bare $d$-band DOS,
which we assume constant for simplicity, and set it as $\Gamma_0=1$ hereafter ) and $q$ (which describes the ratio
between the real and imaginary part of the bare self-energy, and is known as the Fano factor). The
level $\e_0$ is not the bare $f$-level, but rather a renormalized value, which may arise due e.g. to hybridization
with other bands, inelastic scattering effects, interaction renormalization etc. $\e_0$ can be lower than the
original $f$-level \cite{Fogelstrom}, and therefore, a negative value of $\e_0$ does not imply an assumption that the
$f$-level lies below the Fermi energy. The local DOS is given by $\rho=-\frac{1}{\pi} \Im \sum_{k,k'} g_{k,k'} $.
%Apart from the trivial part coming from the bare $d$-band DOS, it has a non-trivial part coming from the interaction
%between the $d$-band and the localized $f$-states. The width of the Fano line-shape is determined by the parameter
%$V^2_0 \Gamma_0=\Gamma_1$. The resulting DOS line-shape is plotted in Fig.~\ref{DOS}(a) (solid line).
 To be specific in comparing our results with experiments we fix parameters
used as $\Gamma_1=V^2_0 \Gamma_0 = 5$ (all energy scales are taken in meV), the bottom of the $f$-band $\e_0=-5$ (shifting this
energy does not alters the line-shape) and the Fano factor $q=1.5$ (these numerical parameters are maintained
hereafter unless stated otherwise). This line-shape is to be compared with that obtained from STM measurements, shown
in the inset of Fig.~\ref{DOS} \cite{Schmidt,Aynajian}.  In our treatment, temperature only enters via the
temperature-dependence of the couplings. The hybridization feature is of $\D \sim 5 $meV$\sim 60K$, and the HO
transition is at $17.5$K, therefore the finite temperature should  smear the DOS line-shape, but not change its features.

The next step is to define the effective interaction between the $f$-electrons and the $d$-holes which gives rise to
the HO. The central part of our proposal  for the HO is to  assume that the {\em most important electron correlations
affect  matrix element  between $d$-band holes with momentum $-Q$ and $f$-electrons with momentum $Q$ (and vice
versa)}, with $Q\approx 0.3$ (the position where the bands cross, Fig.~\ref{bands}). This assumption is consistent
with the observed sharp onset of the hybridization feature at  $Q^*=2Q$ seen in quasiparticle interference \cite{Schmidt}.  Thus, we take the interaction Hamiltonian as
\beq \cH_{I}=\sum_{kk'} U_{kk'} c^\dagger_{k-Q} f^\dagger_{k'-Q} f_{k'+Q} c_{k+Q}+ (Q \to -Q)~~,\eeq where $U_k$ is strongly peaked
around $k=0$ (the width of $U_k$ determines the range of the effective $d-f$ interactions). To continue analytically, we set $k=k'$ and $U_{k}\sim U \delta_{k,0}$,
resulting in the effective Hamiltonian
$ \cH_{I}\approx U_{Q,-Q} c^\dagger_{-Q} f^\dagger_{-Q} f_Q c_{Q}+ (Q \to -Q) \label{Interaction_H}~~.$ The physical meaning of these assumptions is that (i) the most important scattering is with momentum transfer $Q$ \cite{Balatsky,Wiebe,SC}, and (ii) the electron interactions are long-ranged \cite{LongRangeInteractions}. Since the
HO transition seems like a mean-field transition (based on the shape of the heat capacity \cite{Chandra} and the
dependence of the gap on temperature \cite{Aynajian}), it is natural to decouple $\cH_I$ n a mean-field way. We thus
identify the HO order parameter $V$ as
\beq V=V_{HO}=U_{-Q,Q} \langle  c^\dagger_{-Q} f_Q \rangle ~~. \label{OrderParameter} \eeq The 'hidden' nature of this
excitonic order is evident from the
observation that the expectation value of the spin and of the charge (density waves) would be zero in this state hence there is no primary
 single spin and
or charge order in  this hybridization wave state. We do expect a charge density modulation at momentum $Q^*=2Q$ as
a secondary effect, which can be however very small \cite{Su}.
 In the mean-field approximation, $V$ dependens on temperature as $V\propto (T_{HO}-T)^{1/2}$. The interaction hamiltonian in the mean-field approximation takes the form $ \cH_{I,\mathrm{MF}}=V c^\dagger_{-Q} f_Q +h.c. ~$

We now write the Dyson Equation for the Green's function in the presence of $\cH_I$. We point that if $\cH_I$ would
operate on the bare hamiltonian, then no correction (other than at the points $k= \pm Q$) would be observed. However,
the $d$-band is already dressed by the Fano interaction described above, and thus the correction is to all momenta.
Since the bands are symmetric in $k$, we can treat only the $-Q \to Q$ (the full calculation which includes also
$Q\to -Q$ scattering gives similar results). The Dyson equations are
\beqa G_{k,k'} &=& g_{k,k'}+V^2 g_{k,-Q} F_{Q,Q} G_{-Q,k} \nonum
F_{Q,Q} &=& f_{Q,Q}+V^2 f_{Q,Q} G_{-Q,-Q} F_{Q,Q} \label{Dyson2}~~,\eeqa which are easily solved by simple algebra.

%Since the Fano interaction only couples $f_0$ with the $d$-band, we have $f_{Q,Q}=f^{(0)}_Q=\frac{1}{\o-\e_Q-i\eta
%}$.
%Solving for $F$ one obtains $F_{Q,Q}=\frac{1}{\o-\e_Q-V^2 G_{-Q,-Q}}$. Substituting back to Eq.~\ref{Dyson2} and
%solving for $G$, we obtain
%\beqa
%G_{k,k'} &=& g_{k,k'}+\frac{V^2 g_{k,-Q} g_{-Q,k'}}{\o-\e_Q-V^2(g_{-Q,-Q}+G_{-Q,-Q})}\nonum
%G_{-Q,-Q} &=& \frac{1-\sqrt{1-4 V^2 f^{(0)}_Q g_{-Q,-Q}}}{2 V^2 f^{(0)}_Q}~~. \label{Gkk} \eeqa

The resulting local DOS $\rho=\sum_{k,k'} G_{k,k'}$ is plotted in Fig.\ref{DOS}(a) (dashed line) for $V=0.8$ and
$\e_Q=-2.5$. in Fig.~\ref{DOS}(b) we plot the change in DOS due to the HO coupling, $\del \rho=\rho(V)-\rho(V=0)$,
for $V=0.2,0.4,...,1$. As seen, a gap-like feature develops, its width and position depending on the value of $V$
(which in turn depends on temperature). Note also the additional peak at the bottom of the gap, which was observed
experimentally \cite{Schmidt,Aynajian}. For comparison, in the inset of Fig.~\ref{DOS}(b) we plot the experimental
DOS taken from \cite{Schmidt}. The qualitative agreement between the data and theory is
evident. A quantitative fit would require additional parameters to account for the complicated
background DOS seen in the experiments.

\begin{figure}
\includegraphics[width=8truecm]{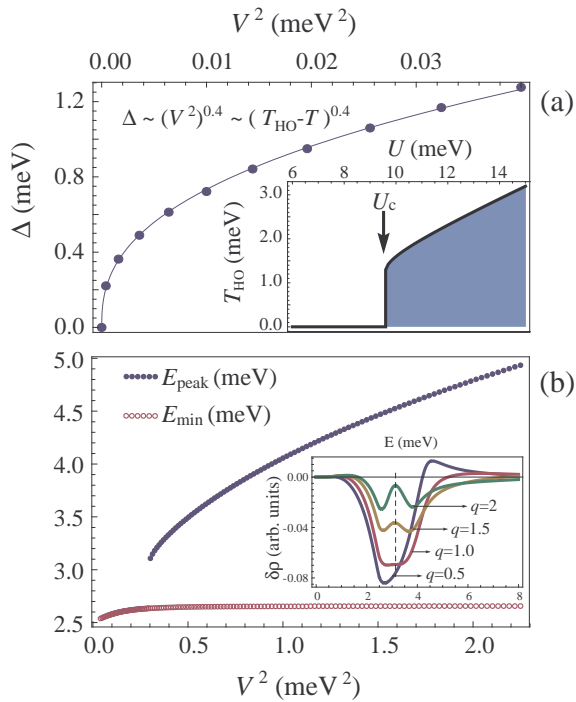}
\caption{ (a) The energy
gap, as obtained from the DOS (Fig.~\ref{DOS}), as a function of the HO order parameter $V$ squared, which is proportional
to $(\Tho-T)$. The solid line is the function
$\D \propto (\Tho-T)^{0.4}$. Inset: the dependence of $T_{HO}$ on the interaction strength $U$, indicating a possible first order transition at $U=U_c$. (b) Position of the minimal point of the DOS
($E_{\mathrm{min}}$, empty circles) and the
position of the peak within the gap ($E_{\mathrm{peak}}$, full circles) as a function of the HO order parameter $V^2$
(which  corresponds to temperature). While $E_{\mathrm{min}}$ saturates at large $V$, $E_{\mathrm{peak}} \sim V \sim
|T -T_{HO}|^{0.4}$  in the same way as the gap $\D$ (Fig.~\ref{Gap}). Inset: change in the DOS, $\del \rho$ for
various values of the Fano factor $q$ (for $V=0.5$).}\label{Gap}
\end{figure}

From the DOS one can extract the size of the energy gap $\D$ as a function of $V$ (for instance, by taking the
distance between the peak just left of the gap and the point where the curvature is maximal just right of the gap).
The dependence of $\D$ on $V^2 \propto (T_{HO}-T)$ is plotted in Fig.~\ref{Gap}(a). We find that $\D\propto
(T_{HO}-T)^{\nu}$, with $\nu\approx 0.4$ (solid line in Fig.~\ref{Gap}(a)), in good agreement with the exponent
extracted from the experimental data \cite{Aynajian} and in agreement with the mean-field treatment assumed.

A pronounced feature of the DOS is the appearance of an additional peak structure within the gap in the DOS
(Fig.~\ref{DOS}(b)). In the experiment \cite{Schmidt,Aynajian} this feature only appears at $T << T_{HO}$ which  we
interpret as large enough $V$, as it grows below $T_{HO}$. In Fig.~\ref{Gap}(b) we have plotted the position of the
minimal point of the DOS ($E_{\mathrm{min}}$, empty circles) and the
position of the peak within the gap ($E_{\mathrm{peak}}$, full circles) as a function of the HO order parameter $V$
(which, in experiment, corresponds to temperature). The peak structure only appears for large enough $V$, and scales
with temperature with the same exponent as the energy gap itself, $E_{\mathrm{peak}} \propto \D \propto
(T_{HO}-T)^{0.4}$. $E_{\mathrm{min}}$, on the other hand, quickly saturates to become temperature-independent. Both
these quantities could be examined in a detailed STM experiment.

In the inset of Fig.~\ref{Gap}(b) we have plotted the DOS change $\del \rho$ for different values of the Fano factor,
$q=0.5,1,1.5,2$ (for $V=0.5$). In the STM experiments, the value of the Fano factor changes, depending on whether the
tip is positioned above a Si site or a U site. Presumably, when the tip is above the Si site it has better coupling
to the $d$-band, which effectively increases the Fano factor. Regardless of the reason for the $q$-factor modulation,
we predict that the height of the peak structure within the gap will also modulate (and will be commensurate with the
modulation of $q$), but its position will not change (dashed line in the inset if Fig.~\ref{Gap}(b)).

The order parameter we discuss here $V_{HO} = U_{-Q,Q}\left\langle c^\dag_{-Q} f_Q\right\rangle$ also carries a
nonzero expectation value for the dipole moment modulation at $Q*$. In the coherent excitonic state $ |\Psi \rangle =
\prod_{k} (u_kc^\dag_{-k} + v_kf^\dag_{k})\left|0\right\rangle $ there is a nonzero expectation value for the dipolar
moment at momentum $Q*$: $ \left\langle | P_{2Q} |\right\rangle = \left\langle \Psi| \hat{r} e^{i2Q r}
|\Psi\right\rangle $,  as long as the parity of $d$-and $f$-orbitals is different. This modulation of the dipole
moment could be observed with probes that couple to electric density modulation, for example an STM with a
ferroelectric (or hybrid ferroelectric-metallic) tip.

The formalism described above allows us to calculate the critical temperature of the HO transition. Working in a Nambu-like space mixing $d$- and $f$-electrons and using Eq.~\ref{OrderParameter}, we
identify the self-consistency equation for the HO,
%\beq
%-\frac{1}{U_{-Q,Q}}=T \sum_{i \omega_n} \frac{1}{f_Q(i \omega_n)^{-1}g_{-Q}(i \omega_n)^{-1}-V_{HO}^2}
%~~,\label{Self-consistency equation}\eeq where $f_Q(i \omega_n)$ and $g_{-Q}(i \omega_n)$ are the Matsubara Green's
%function in the presence of the Fano interaction, evaluated at momentum $Q$, and $i \omega_n$ are matsubara
%frequencies. The resulting $V_{HO}$ clearly has a $(T_{HO}-T)^{1/2}$ dependence close to the transition. At the
%transition, $T=T_{HO}$ and $V_{HO}=0$, and therefore $T_{HO}$ is defined by the equation
\beq -\frac{1}{U_{-Q,Q}}=T_{HO} \sum_{i \omega_n} f_Q(i \omega_n,Q)g_{-Q}(i \omega_n) \label{T_HO equation} ~~,\eeq where $f_Q(i \omega_n)$ and $g_{-Q}(i \omega_n)$ are the Matsubara Green's
function in the presence of the Fano interaction, evaluated at momentum $Q$, and $i \omega_n$ are Matsubara
frequencies.
We have calculated the critical temperature $T_{HO}$ for the parameters above as a function of $U$, shown in the
inset of Fig.~\ref{Gap}(a). We find that for $U$ below a certain value $U_c$, Eq.~\ref{T_HO equation} does not have a
solution, due to the compact nature of the interaction in momentum space. This feature may explain why the hidden
order has not been universally observed in other heavy-Fermion materials, even ones with a similar band structure. It may also be relevant to the destruction of the HO and appearance of magnetic ordering with applied pressure, with a 1st order transition between them \cite{Hassinger, Niklowitz}
A possible scenario for the connection between the HO and  magnetic order (clearly avoided in the present study which discusses spinless electrons) is the following. As pressure is applied the interaction strength decreases, until finally it becomes lower than $U_c$. The magnetic interaction between the localized $f$-electrons, previously screened by the conduction electron which are coupled to the $f$-electrons in the HO state, becomes significant and gives rise to magnetic ordering. This scenario will be studied in detail in the future.

We thank P. W\"{o}lfle, M. Graf, A. Chantis, K. Haule, T. Durakiewicz, G. Kotliar and J.-X. Zhu for valuable discussions, and A.
Schmidt, M. Hamidian and C. S. Davis for stimulating discussions and for sharing with us their STM data.
  This work was supported by   BES, UCOP-TR-01 and  in part,  by grant No. DE-AC52-06NA25396 and

{}

\begin{thebibliography}{}
% Hidden Order
\bibitem{Palstra}
T. T. M. Palstra \etal, \prl {\bf 55},
2727 (1985).

\bibitem{Schlabitz}
W. Schlabitz \etal,  Z. Phys. B {\bf 62}, 171 (1986).

\bibitem{Maple}
M. B. Maple \etal,  Phys. Rev. Lett.
{\bf 56}, 185 (1986).

\bibitem{Tripathi}
V. Tripathi \etal, Nature Physics {\bf 3}, 78 (2007).

\bibitem{Shah}
N. Shah \etal, \prb {\bf 61}, 564 (2000).

% Hidden Order as localized transition: Chandra,Santini,Kasuya,Sikkema,Cricchio,Haule
\bibitem{Chandra}
P. Chandra \etal, Nature {\bf 417}, 831 (2002).

 \bibitem{Santini}
P. Santini, Phys. Rev. B {\bf 57}, 5191 (1998).

\bibitem{Kasuya}
T. Kasuya, J. Phys. Soc. Jpn. {\bf 66}, 3348 (1997).

\bibitem{Sikkema}
 A. E. Sikkema, \etal, Phys. Rev. B {\bf 54}, 9322 (1996).

\bibitem{Cricchio}
F. Cricchio \etal, Phys. Rev. Lett. {\bf 103}, 107202 (2009).

\bibitem{Haule}
K. Haule and G. Kotliar, Nature Physics {\bf 5}, 796 (2009).

% Hidden Order as Itinerant transition: Ikeda,Varma,Balatsky
\bibitem{Ramirez}
 A. P. Ramirez \etal,  Phys. Rev. Lett. {\bf  68}, 2680 (1992).

\bibitem{Ikeda}
H. Ikeda and Y. Ohashi, Phys. Rev. Lett. {\bf 81}, 3723 (1998).

\bibitem{Varma}
C. M. Varma and L. Zhu,  Phys. Rev. Lett. {\bf 96}, 036405 (2006).

\bibitem{Elgazzar}
S. Elgazzar \etal, Nature Materials {\bf 8}, 337 (2009).

\bibitem{Harima}
H. Harima \etal, J. Phys. Soc. Jpn. {\bf 79}, 033705  (2010).

\bibitem{Balatsky}
A. V. Balatsky \etal, Phys. Rev. B {\bf 79}, 214413 (2009).

%Experiments
\bibitem{Wiebe}
C. R. Wiebe \etal, Nature physics {\bf 3}, 96 (2007).

\bibitem{Broholm}
C. Broholm, Phys. Rev. Lett. {\bf 58}, 1467 (1987).

\bibitem{Riseborough}
for recent reviews see e.g., P. S. Riseborough, Adv. Phys. {\bf 49}, 257 (2000);
D. T. Adroja \etal, J. Opt. Adv. Mat. {bf 10}, 1564 (2008).

%\bibitem{Adroja}
%D. T. Adroja \etal, J. Opt. Adv. Mat. {bf 10}, 1719 (2008).

\bibitem{Santander}
A. F. Santander-Syro, Nature physics {\bf 5}, 637 (2009).

\bibitem{Schmidt}
A. R. Schmidt, Nature {\bf 465}, 570 (2010).

\bibitem{Aynajian}
P. Aynajian \etal, PNAS {\bf 107}, 10383 (2010).

%\bibitem{Chantis}
%Chantis, A., Dubi, Y. and Balatsky, A. V., unpublished.
\bibitem{Madhavan}
V. Madhavan \etal, Phys. Rev. B {\bf 64}, 165412 (2001).

\bibitem{Yang}
Y. -F. Yang, \prb {\bf 79}, 241107
(2009).

\bibitem{Maltseva}
M. Maltseva, M. Dzero and P. Coleman, \prl {\bf 103}, 206402 (2009).

\bibitem{Figgins}
J. Figgins and D. K. Morr, Phys. Rev. Lett. {\bf 104}, 187202 (2010).

\bibitem{Fogelstrom}
M. Fogelstr\"{o}m \etal, Phys. Rev. B {\bf 82}, 014527 (2010).

\bibitem{Wolfle}
P. Wolfle, Y. Dubi and A. V. Balatsky, Phys. Rev. Lett. {\bf 105}, 246401 (2010).

\bibitem{Cheng}
P. Cheng \etal, Phys. Rev. Lett. {\bf 105}, 076801 (2010), supplementary material at cond-mat/1001.3220.

\bibitem{SC}
This assumption is consistent with the appearance of superconductivity at lower temperatures, since it implies the
the Fermi surface is only partially gapped, allowing for SC correlations to develop.

\bibitem{LongRangeInteractions}
The fact that the interaction is limited to $k=k'=0$ implies infinite-long-ranged electron interactions. While
this is clearly an approximation (taken to allow for an analytic solution) the fact that the interactions might not be short-ranged is quite reasonable, noting that the screening length in this system could be reasonably long.

\bibitem{Su}
J.-J. Su \etal,  J. Phys. Condens. Matter, in press.

\bibitem{Hassinger}
E. Hassinger \etal, \prb {\bf 77}, 115117 (2008).

\bibitem{Niklowitz}
P. G. Niklowitz \etal, \prl {\bf 104}, 106406 (2010).


\end{thebibliography}
\end{document}